\documentclass[12pt]{article}
\usepackage{amssymb,amsmath,cite,mathdots}

\def\beq{\begin{equation}}
\def\eeq{\end{equation}}
\def\bea{\begin{eqnarray*}}
\def\eea{\end{eqnarray*}}
\def\nn{\nonumber}

\def\bra#1{\left\langle #1\right|}
\def\ket#1{\left| #1\right\rangle}
\def\bracket#1#2{\left\langle #1 | #2 \right\rangle}

\def\N0{ {\mathbb Z}_{+} }
\def\v0{ |d,r) }

\def\proof#1{{\bf Proof:} #1 $\blacksquare$ \medskip}

\newtheorem{lemma}{Lemma}
\newtheorem{prop}[lemma]{Proposition}
\newtheorem{thm}{Theorem}

\begin{document}

\begin{center}
{\large\bf Matrix elements and duality for type 2 unitary representations of the Lie superalgebra $gl(m|n)$}\\
~~\\

{\large Jason L. Werry, Mark D. Gould and Phillip S. Isaac }\\
~~\\

School of Mathematics and Physics, The University of Queensland, St Lucia QLD 4072, Australia.
\end{center}

\begin{abstract}
The characteristic identity formalism discussed in our recent articles is further utilized
 to derive matrix elements of type 2 unitary irreducible $gl(m|n)$ modules. In particular, we give matrix element formulae for
all $gl(m|n)$ generators, including the non-elementary generators, together with their phases on finite dimensional
type 2 unitary irreducible representations. Remarkably, we find that the type 2 unitary matrix element equations coincide with the type 1 unitary matrix element equations for non-vanishing matrix elements up to a phase.
\end{abstract}

%

\section{Introduction}
 
This is the third paper in a series aimed at deriving matrix elements of elementary and
non-elementary generators of finite dimensional unitary irreducible representations for the
Lie superalgebra $gl(m|n)$. The concept of a conjugation operation (necessary to understand
unitary representations) was developed by Scheunert, Nahm and Rittenberg \cite{SNR1977}. 
These unitary representations were then classified in the work of Gould and Zhang \cite{ZhaGou1990,GouZha1990}. 
The above work shows that there are two types of finite dimensional irreducible unitary 
representations of $gl(m|n)$ that are defined depending on the sesquilinear form that exists
on the module. In this paper, we consider the generator matrix elements of 
irreducible type 2 unitary representations which up to now have not featured in the literature.  

In the first paper of this series \cite{GIW1} we constructed invariants associated
with $gl(m|n)$, and obtained analytic expressions for their eigenvalues. 
The second paper in this series \cite{GIW2} utilized these results to obtain matrix 
elements for the irreducible type 1 unitary representations. A goal of this series 
of papers has been to highlight the innovative techniques involving characteristic 
identities \cite{Green1971,BraGre1971,OBCantCar1977,Gould1985}. Characteristic 
identities associated to Lie superalgebras have been studied in the work of 
Green and Jarvis \cite{JarGre1979,GreJar1983} and Gould \cite{Gould1987}. We expect the 
utility and importance of these characteristic identities will become increasingly 
evident as this series continues. For a detailed survey of the literature on the subject, 
and for a broad setting of the current work, we direct the reader to the initial article 
in this series \cite{GIW1}. 

The highest weight of a unitary $gl(m|n)$ module is related to the highest weight of its dual in a non-trivial manner relative 
to the $gl(m)$ case. In general, taking the dual of a $gl(m|n)$ module involves a combinatorial procedure as opposed
to an algebraic one and is directly related to the atypicality of the module in question. In this paper we investigate
this duality and show how the additional branching rules required for unitary $gl(m|n)$ modules appear more natural when
consistency under duality is considered.

The paper is organised as follows. Section \ref{prelim} provides a brief review of the context
and important notations used throughout the paper. In Section \ref{ContravariantReps}   
we present the three main subclasses of type 2 unitary representations that are under 
consideration. After giving details of the type 2 unitary branching rules 
in Section \ref{branch} we then investigate the behavior of the branching rules
under duality in Section \ref{duality}. Finally, we give a construction of the explicit matrix element
formulae in Section \ref{MEs}.   


\section{Preliminaries}
\label{prelim}

We continue the same notation as used in the previous articles of this series \cite{GIW1,GIW2} 
which we summarize here for convenience. The graded index notation requires Latin 
indices $1\leq i,j,\ldots\leq m$ to be assumed even, and Greek indices
$1\leq\mu,\nu,\ldots\leq n$ taken to be odd. The parity of the index is given by
$$
(i)=0,\ \ (\mu)=1.
$$
Where convenient we may use ungraded indices $1\leq p,q,r,s\leq m+n$. 
For indices in the range $p=1,\ldots,m$ we have the parity
$(p)=0$, and for indices $p=m+\mu$ for some $\mu=1,\ldots,n$ the parity is $(p)=(\mu)=1.$ 
The $gl(m|n)$ generators $E_{pq}$ satisfy the graded commutation relations
$$
[E_{pq},E_{rs}] = \delta_{qr}E_{ps} - (-1)^{[(p)+(q)][(r)+(s)]}\delta_{ps}E_{rq}.
$$

The Cartan subalgebra is given by the set of mutually commuting generators $E_{aa}$ 
whose eigenvalues label the weights occurring in a given $gl(m|n)$ module. A weight may be expanded in terms
of the fundamental weights $\varepsilon_i$ ($1\leq i\leq m$) and $\delta_\mu$ ($1\leq \mu
\leq n$) \cite{Kac1977}. These fundamental weights provide a basis for $H^*$. 
We may therefore expand a weight $\Lambda\in H^*$ as
$$
\Lambda = \sum_{i=1}^m\Lambda_i\varepsilon_i + \sum_{\mu=1}^n\Lambda_\mu\delta_\mu.
$$
In this basis, the root system is given by the set of even roots
\begin{align}
\pm(\varepsilon_i-\varepsilon_j), & \ \ 1\leq i<j\leq m,\nn\\
\pm(\delta_\mu - \delta_\nu), & \ \ 1\leq \mu<\nu\leq n,\nn
\end{align}
and the set of odd roots
\begin{align}
\pm(\varepsilon_i-\delta_\mu), \ \ 1\leq i\leq m,\ \ 1\leq \mu\leq n.
\end{align}
A system of simple roots is given by the distinguished set
\begin{equation}
\left\{ \left. \varepsilon_i-\varepsilon_{i+1},\ \varepsilon_m-\delta_1,\
\delta_\mu-\delta_{\mu+1}\ \right| \ 1\leq i<m,\ 1\leq \mu<n \right\}.
\label{SimpleRoots}
\end{equation}
The sets of even and odd positive roots are then given, respectively, by
\begin{align}
\Phi_0^+ &= \{ \varepsilon_i - \varepsilon_j\ |\ 1\leq i<j\leq m\} \cup \{\delta_\mu -
\delta_\nu\ |\ 1\leq \mu<\nu\leq n\},\nn\\
\Phi_1^+ &= \{ \varepsilon_i - \delta_\mu\ |\ 1\leq i\leq m,\ 1\leq\mu\leq n\}. \nn
\end{align}
An important quantity is the graded half-sum of positive roots defined by 
\begin{align}
\rho &= \frac12 \sum_{\alpha\in\Phi_0^+}\alpha - \frac12\sum_{\beta\in\Phi_1^+}\beta\nn\\
&= \frac12 \sum_{j=1}^m(m-n-2j+1)\varepsilon_j +
\frac12\sum_{\nu=1}^n(m+n-2\nu+1)\delta_\nu.
\label{rho}
\end{align}


Every finite dimensional irreducible $gl(m|n)$ module is a 
$\mathbb{Z}_2$-graded vector space
$$
V=V_0\oplus V_1,
$$
(so that $v\in V_j$ implies the grading $(v)=j$ for $j=0,1$)
which admits a highest weight vector, whose
weight $\Lambda$ uniquely characterizes the representation. We denote the corresponding irreducible
highest weight module by $V(\Lambda)$ and the associated representation by
$\pi_\Lambda$. Relative to the $\mathbb{Z}_2$-grading, it is
assumed, unless stated otherwise, that the highest weight vector $v^\Lambda$ has an even
grading, i.e. $v^\Lambda\in V(\Lambda)_0$. As a simple example, the fundamental vector representation is denoted $V(\varepsilon_1)$ using this
notation. 


Components of the highest weight $\Lambda$ satisfy the lexicality conditions
$$
\Lambda_i - \Lambda_j\in \mathbb{Z}_+\ (1\leq i<j\leq m),\ \ \Lambda_\mu-\Lambda_\nu\in
\mathbb{Z}_+ \ (1\leq \mu<\nu\leq n).
$$
We refer to such a weight as dominant.

{\bf Note:} While the components of a dominant weight $\Lambda$ must satisfy the above
lexicality conditions we note that $(\Lambda,\epsilon_m - \delta_1)$ may be any complex number.
This gives rise to a 1-parameter families of finite dimensional irreducible modules.

The fundamental vector representation $\pi_{\varepsilon_1}$ of $gl(m|n)$ is $m+n$ dimensional with a
basis $\{ \ket{a}\ |\ 1\leq a\leq m+n\}$ on which the generators $E_{ab}$ have the
following action:
$$
E_{ab}\ket{d} = \delta_{bd}\ket{a},
$$
so that
$$
\bra{c}E_{ab}\ket{d} = \delta_{bd}\bracket{c}{a} = \delta_{bd}\delta_{ac}
$$
or alternatively
$$
\pi_{\varepsilon_1}\left( E_{ab} \right)_{cd} = \delta_{bd}\delta_{ac}.
$$
This gives rise to a non-degenerate even invariant bilinear form on $gl(m|n)$ defined by
$$
(x,y) = \mbox{str}(\pi_{\varepsilon_1}(xy)) = \sum_{i=1}^m\pi_{\varepsilon_1}(xy)_{ii}
-\sum_{\mu=1}^n\pi_{\varepsilon_1}(xy)_{\mu\mu}.
$$
In particular we have 
\begin{eqnarray}
\left( E_{ab},E_{cd} \right) 
& = & (-1)^{(d)}\delta_{bc}\delta_{ad},
\label{eq2}
\end{eqnarray}
which leads to a bilinear form on the fundamental weights
$$
(\varepsilon_i,\varepsilon_j) = \delta_{ij},\ \ (\varepsilon_i,\delta_\mu)=0, \ \ (\delta_\mu,\delta_\nu) =
-\delta_{\mu\nu},
$$
which in turn induces a non-degenerate bilinear form on our weights $\Lambda$ given by
\begin{equation}
\left( \Lambda,\Lambda'\right) = \sum_{i=1}^m\Lambda_i\Lambda_i' -
\sum_{\mu=1}^n\Lambda_\mu\Lambda_\mu'.
\label{form}
\end{equation}
 
On every irreducible, finite dimensional $gl(m|n)$-module $V(\Lambda)$, there exists a 
sesquilinear form $\langle\ |\ \rangle_\theta$ with the distinguished property \cite{ZhaGou1990,GouZha1990} 
$$
\langle E_{pq}v | w\rangle_\theta = (-1)^{(\theta - 1)[(p)+(q)]}\langle v|E_{qp} w\rangle_\theta,
$$
with $\theta = 1$ or 2 relating to two inequivalent forms. The irreducible, finite
dimensional module $V(\Lambda)$ is said to be type $\theta$ unitary if $\langle\ |\
\rangle_\theta$ is positive definite on $V(\Lambda)$, and the corresponding
representation is said to be type $\theta$ unitary. Equivalently, for a finite dimensional unitary
irreducible representation $\pi$, we require that the
linear operators $\pi(E_{pq})$ satisfy 
\begin{equation}
\left[ \pi(E_{pq}) \right]^\dagger = (-1)^{(\theta - 1)[(p)+(q)]} \pi(E_{pq}),
\label{hermit}
\end{equation}
where $\dagger$ denotes the usual Hermitian conjugation such that
$$
\left( \left[ \pi(E_{pq}) \right]^\dagger \right)_{\alpha\beta} = {\left[
\overline{\pi(E_{pq})} \right]}_{\beta\alpha},
$$
with $\overline{A}$ denoting the matrix with complex entries conjugate to those of $A$.

Given a representation $\pi$, its dual representation $\pi^*$ is defined by
\cite{NahmSch1976} 
$$
\pi^*(E_{pq}) = -\left[\pi(E_{pq})\right]^T,
$$
where $T$ denotes the supertranspose. On a homogeneous basis 
$\{e_\alpha\}$ of $V$, the supertranspose is defined as
$$
\left( \left[\pi(E_{pq})\right]^T\right)_{\alpha\beta} = (-1)^{[(p)+(q)](\beta)}
\left[\pi(E_{pq})\right]_{\beta\alpha},
$$
where $(\beta)$ denotes the grading of basis vector $e_\beta$.

It was shown in \cite{ZhaGou1990,GouZha1990} that both type 1 and 2 unitary irreducible representations 
are completely characterized by conditions on the highest weight labels. 
This classification is given by the following three theorems.
\begin{thm} \label{gzthm1}
The irreducible highest weight $gl(m|n)$-module $V(\Lambda)$ is type 1 unitary if and only
if $\Lambda$ is real and satisfies
\begin{itemize}
\item[(i)] $(\Lambda+\rho,\varepsilon_m - \delta_n)>0;$ or
\item[(ii)] there exists an odd index $\mu\in\{1,2,\ldots,n\}$ such that
\begin{equation} \label{EqnType1Atyp}
(\Lambda+\rho,\varepsilon_m-\delta_\mu)=0 = (\Lambda,\delta_\mu-\delta_n).
\end{equation}
\end{itemize}
\end{thm}

\begin{thm} \cite{GouZha19902}
The dual of a type 1 unitary irreducible representation is a type 2 unitary representation
and vice versa.
\end{thm}

\begin{thm} \cite{ZhaGou1990,GouZha1990}
\label{gzthm2}
The irreducible highest weight $gl(m|n)$-module $V(\Lambda)$ is type 2 unitary if and only
if $\Lambda$ is real and satisfies
\begin{itemize}
\item[(i)] $(\Lambda+\rho,\varepsilon_1 - \delta_1)<0;$ or
\item[(ii)] there exists an even index $k\in\{1,2,\ldots,m\}$ such that
\begin{equation} \label{EqnType2Atyp}
(\Lambda+\rho,\varepsilon_k-\delta_1)=0 = (\Lambda,\varepsilon_k-\varepsilon_1).
\end{equation}
\end{itemize}
\end{thm}

When considering dual modules we shall make direct use of proposition 5 given in \cite{ZhaGou1990} which we give here for convenience:

\begin{prop}
\label{DualWeight}
Consider a type 1 module $V(\Lambda)$. If $\Lambda$ is atypical we set $\mu$ equal to the odd index that satisfies (\ref{EqnType1Atyp}). Otherwise we set $\mu = n + 1$. 

Define a sequence of odd indices $\mu_i, 1 \leq i \leq m$, by
$$
\mu_i = [\mu_m  + (\Lambda, \epsilon_i - \epsilon_m)] \wedge n, ~~a \wedge b = min(a,b), 
$$
where 
$$
\mu_m = \mu - 1.
$$
Then,\\ 
(i) the highest weight of the minimal $\mathbb{Z}$-graded component
of the irreducible $gl(m|n)$ module $V(\Lambda)$ is
$$
\bar{\Lambda} = \Lambda - \sum_{i=1}^m \sum_{\nu=1}^{\mu_i} (\epsilon_i - \delta_\nu);
$$\\
(ii) the lowest weight of $V(\Lambda)$ is
$$
\Lambda^- = \tau(\bar{\Lambda}),
$$
where $\tau$ is the unique Weyl group element sending the positive
even roots into negative ones;\\
(iii) $V(\Lambda)$ admits $d_\Lambda + 1$ levels with 
$$
d_\Lambda = \sum_{i=1}^m \mu_i ;
$$\\
(iv) The highest weight of the dual module $V^*(\Lambda)$ is 
$$
\Lambda^* = - \Lambda^-
$$
\end{prop}

\section{Characterisation of contravariant tensor and non-tensorial representations}
\label{ContravariantReps}

We now adopt an approach similar to that presented in the article \cite{GIW2}, by outlining a straightforward
characterisation of the type 2 unitary representations of $gl(m|n)$. For this case, we introduce the
system $\overline{\Phi}'$ of extended simple roots:
\begin{align*}
\overline{\varphi}_1 &= -\varepsilon_1,\\
\overline{\varphi}_i &= \varepsilon_{i-1} - \varepsilon_i,\ \ 1<i\leq m,\\
\overline{\varphi}_{\bar{1}} &= \varepsilon_m - \delta_1,\\
\overline{\varphi}_\mu &= \delta_{\mu-1}-\delta_\mu,\ \ 1<\mu\leq n.
\end{align*}
Here we have extended the set of simple roots given in (\ref{SimpleRoots}) by including the
additional weight $\overline{\varphi}_1.$ We also remark that we use the ``overbar'' notation to
indicate that $\overline{\Phi}'$ makes use of an extension different to that introduced in
\cite{GIW2} for the type 1 unitary case. 
We also use the notation $\bar{1}$ to indicate an odd index (i.e. $\mu=1$ in this case). 

We may define a weight basis dual (in the graded sense) to $\overline{\Phi}'$ with respect to the
form (\ref{form}) as follows:
\begin{align*}
\overline{\omega}_i &=
(\underbrace{0,0,\ldots,0}_{i-1},\underbrace{-1,-1,\ldots,-1}_{m-i+1}|\underbrace{1,1,\ldots,1}_n),\
\ 1\leq i\leq m,
\\
\overline{\omega}_\mu &=
(\underbrace{0,0,\ldots,0}_m|\underbrace{0,0,\ldots,0}_{\mu-1},\underbrace{-1,-1,\ldots,-1}_{n-\mu+1}),
\ 1\leq \mu\leq n.
\end{align*}
These are analogous to the fundamental dominant weights for Lie algebras.
Explicitly we have
$$
(\overline{\omega}_i,\overline{\varphi}_j) = \delta_{ij},\ \
(\overline{\omega}_\mu,\overline{\varphi}_\nu) = -\delta_{\mu \nu}, \ \
(\overline{\omega}_i,\overline{\varphi}_\nu) = 0 = (\overline{\omega}_\nu,\overline{\varphi}_i).
$$

Based on the classification theorems of unitary representations of $gl(m|n)$ given in
\cite{GouZha1990,ZhaGou1990}, we make the observation that for $1\leq i\leq m$, the
$\overline{\omega}_i$ correspond to type 1 unitary dominant weights, and for $1\leq \mu\leq n$, the
$\overline{\omega}_\mu$ correspond to type 2 unitary dominant weights.\footnote{Actually,
$\overline{\omega}_1$ gives a one-dimensional highest weight corresponding to both a type 1 and 2
unitary representation. In fact, any real multiple of this weight will give rise to a type 1 and 2
highest weight representation. See Lemma 2 of \cite{GIW2} and the comment immediately following its
proof.}

Using similar arguments given in \cite{GIW2} for the type 1 unitary case, we may state a Theorem
which is the analogue of Theorem 2 from \cite{GIW2} for the type 2 unitary case.

\begin{thm} \label{thmtp}
Let $V(\Lambda)$ and $V(\Lambda')$ be irreducible type 2 unitary modules. Then $V(\Lambda +
\Lambda')$ is also irreducible type 2 unitary and occurs in $V(\Lambda)\otimes V(\Lambda')$.
\end{thm}

It is clear that we may use the fundamental dominant weight analogues given above to expand any highest weight
$\Lambda$ as
\begin{equation}
\Lambda = \sum_{i=1}^m(\Lambda,\overline{\varphi}_i)\overline{\omega}_i -
\sum_{\mu=1}^n(\Lambda,\overline{\varphi}_\mu)\overline{\omega}_\mu.
\label{LambdaExpansion}
\end{equation}
Using this expansion, however, it is not apparent after applying the result of Theorem \ref{thmtp} whether or not the module
$V(\Lambda)$ is type 2 unitary. We instead describe the weights in terms of a slightly modified set, which we refer
to as the {\em type 2 unitary graded fundamental weights}, defined as
\begin{align*}
\delta &= \overline{\omega}_1,\\
\overline{\Omega}_i &= \overline{\omega}_i + (m-i+1)\overline{\omega}_{\bar{1}},\ \ 1<i\leq m,\\
\overline{\varepsilon} &= \overline{\omega}_{\bar{1}},\\
\overline{\Omega}_\mu &= \overline{\omega}_\mu, \ \ 1<\mu \leq n.
\end{align*}
Using these weights, we may rewrite the expansion (\ref{LambdaExpansion}) as 
\begin{equation}
\Lambda = \sum_{i=2}^m(\Lambda,\overline{\varphi}_i)\overline{\Omega}_i -
\sum_{\mu=2}^n(\Lambda,\overline{\varphi}_\mu)\overline{\Omega}_\mu
-\left( (\Lambda,\overline{\varphi}_{\bar{1}} ) + \sum_{i=2}^m(m-i+1)(\Lambda,\overline{\varphi}_i) \right) \overline{\varepsilon}
+(\Lambda,\overline{\varphi}_1)\delta.
\label{FDWExpansion}
\end{equation}
Note that the coefficients $(\Lambda,\overline{\varphi}_i)$ and $-(\Lambda,\overline{\varphi}_\mu)$
are always positive integers, and so by the result of Theorem \ref{thmtp} these terms shall always
contribute to irreducible type 2 representations that are contravariant tensorial.

The coefficient of $\overline{\varepsilon}$ in (\ref{FDWExpansion}) may in some cases be negative.
We may combine part of this coefficient with the first
two terms to contribute to an overall unitary type 2 contravariant tensorial representation. What is left over is
characterised by the result of the following Lemma.

\begin{lemma} \label{gamlem}
The irreducible $gl(m|n)$ module $V(\gamma\overline{\varepsilon})$ is type 2 unitary if and only
if $\gamma=0,1,2,\ldots,m-1$ or $m-1<\gamma\in\mathbb{R}$.
\end{lemma}

\proof{
The proof follows immediately from the classification of Theorem \ref{gzthm2}. When
$\gamma=0,1,2,\ldots,m-1,$ $V(\gamma\overline{\varepsilon})$ will be atypical, otherwise for
$m-1<\gamma\in\mathbb{R}$, $V(\gamma\overline{\varepsilon})$ is typical.
}

Note that when $\gamma$ takes on integer values, even with $m-1<\gamma\in\mathbb{Z}$,
$V(\gamma\overline{\varepsilon})$ will determine a contravariant tensor representation. For
noninteger values of $\gamma$, the ensuing representation will be nontensorial.

As we have already remarked, for any $\omega\in\mathbb{R}$, the module $V(\omega \delta)$ is type 2
unitary (also type 1 unitary) and one-dimensional. This is the only subclass of unitary module that
can be taken as either type 1 or type 2 unitary.

In summary, we have the following result.

\begin{thm}
The highest weight $\Lambda$ of an irreducible type 2 unitary $gl(m|n)$ representation is
expressible as
$$
\Lambda = \Lambda_0 + \gamma \overline{\varepsilon} + \omega \delta,
$$  
where $\Lambda_0$ is the highest weight of an irreducible contravariant tensorial (type 2 unitary)
representation, $\gamma\in \mathbb{R}$ satisfies the conditions of Lemma \ref{gamlem}, and
$\omega\in\mathbb{R}$. 
\end{thm}

The key point is that since $V(\Lambda_0)$, $V(\gamma\overline{\varepsilon})$ and $V(\omega\delta)$
are all type 2 unitary representations, by Theorem \ref{thmtp}, any type 2 unitary module will occur
in the tensor product of these three. In this sense, we have identified that the contravariant tensor
modules, the modules $V(\gamma\overline{\varepsilon})$ and the one-dimensional modules
$V(\omega\delta)$ are the building blocks for type 2 unitary modules.

\section{Branching rules} 
\label{branch}
In this section we will now obtain the $gl(m|n+1)\downarrow gl(m|n)$ branching rules for type 2 unitary modules. 
Let $\lambda_{r,p}$ be the weight label located at the $r$th position in the $p$th row of the GT pattern for $gl(m|n+1)$ that is written as 
\begin{equation}
\left|
\begin{array}{cccccccccc}
\lambda_{1,m+n+1} & \lambda_{2,m+n+1} & \cdots & \lambda_{m,m+n+1} & | & 
\lambda_{\bar{1},m+n+1} & \lambda_{\bar{2},m+n+1} & \cdots &
\lambda_{\bar{n},m+n+1} & \lambda_{\overline{n+1},m+n+1} 
\\
\lambda_{1,m+n} & \lambda_{2,m+n} & \cdots & \lambda_{m,m+n} & | &
\lambda_{\bar{1},m+n} & \lambda_{\bar{2},m+n} &\cdots & \lambda_{\bar{n},m+n} &\\
\vdots &  & & & \vdots & &   & \iddots && \\
\lambda_{1,m+1} & \lambda_{2,m+1} & \cdots & \lambda_{m,m+1} & | & \lambda_{\bar{1},m+1} &&&&\\
 &  & \cdots &  &  &  &  &  &  &   \\
\lambda_{1,m} & \lambda_{2,m}& \cdots & \lambda_{m,m} &  &&&&&\\
\vdots &  & \iddots & &  & &  & && \\
\lambda_{1,2} & \lambda_{2,2} & &  & &  & & & &\\
\lambda_{1,1} & & & & & & & & &  
\end{array}
\right)
\label{genGT}
\end{equation}
and where each row is a highest weight corresponding to an irreducible representation
permitted by the branching rule for the subalgebra chain
\begin{equation}
gl(m|n+1)\supset gl(m|n)\supset \cdots \supset gl(m|1)\supset gl(m)\supset
gl(m-1)\supset \cdots \supset gl(1).
\label{flag}
\end{equation}

Using the notation above we first recall the {\em branching conditions}
given in \cite{GIW1}, which provide necessary conditions on the $gl(m|p)$ highest
weights occurring in the branching rule of an irreducible $gl(m|p+1)$ highest weight
representation.

\begin{thm} \label{pg1} \cite{GIW1}
For $r\geq m+1$, the following conditions on the dominant weight labels must hold in the pattern
(\ref{genGT}):
\begin{align*}
\lambda_{\mu,r+1}\geq \lambda_{\mu,r}\geq \lambda_{\mu+1,r+1}, & \ \ 1\leq \mu\leq n,\\
\lambda_{i,r+1} \geq \lambda_{i,r}\geq \lambda_{i,r+1}-1, &\ \ 1\leq i\leq m.
\end{align*}
\end{thm}
The results of \cite{GouBraHug1989,GouJarBra1990} provide stronger conditions for the case
$gl(m|1)\supset gl(m)$:

\begin{thm} \label{pg2} \cite{GouBraHug1989,GouJarBra1990}
For a unitary type 2 irreducible module $V(\Lambda)$ of $gl(m|1)$, 
using the notation of (\ref{genGT}),
we have the following conditions on the dominant weight labels: 
\begin{align*}
\lambda_{i,m+1}\geq \lambda_{i,m}\geq \lambda_{i,m+1}-1, &\ \
\mbox{if }(\Lambda+\rho,\varepsilon_i-\delta_1)<0 \mbox{ (i.e. only if $\Lambda$ typical),}\\
\lambda_{i,m}= \lambda_{i,m+1}, &\ \
\mbox{if }(\Lambda+\rho,\varepsilon_i-\delta_1)=0 \mbox{ (i.e. only if $\Lambda$
atypical)}\\
\end{align*}
while for unitary type 1 irreducible representations 
\begin{align*}
\lambda_{i,m+1} \geq \lambda_{i,m}\geq \lambda_{i,m+1}-1, &\ \ 1\leq i\leq m-1, \\
\lambda_{m,m+1}\geq \lambda_{m,m}\geq \lambda_{m,m+1}-1, &\ \
\mbox{if }(\Lambda+\rho,\varepsilon_m-\delta_1)<0 \mbox{ (i.e. only if $\Lambda$ typical),}\\
\lambda_{m,m}= \lambda_{m,m+1}, &\ \
\mbox{if }(\Lambda+\rho,\varepsilon_m-\delta_1)=0 \mbox{ (i.e. only if $\Lambda$
atypical).}\\
\end{align*}
\end{thm}

For the general $gl(m|n+1)$ branching rule, we have the following result.
\begin{thm} \label{mainbranchingrule}
For a unitary type 2 irreducible $gl(m|n+1)$ representation, the basis vectors can be
expressed in the form (\ref{genGT}), with the following conditions on the dominant weight labels:
\begin{itemize}
\item[(1)] 
For $r\geq m+1$, \\
$\lambda_{\mu,r+1}\geq \lambda_{\mu,r}\geq \lambda_{\mu+1,r+1},$ $1\leq \mu\leq n, $\\ 
$\lambda_{i,r+1} \geq \lambda_{i,r}\geq \lambda_{i,r+1}-1,$ $1\leq i\leq m$, \\
(i.e result of Theorem \ref{pg1});
\item[(2)] 
$\lambda_{i,m+1}\geq \lambda_{i,m}\geq \lambda_{i,m+1}-1,$
if $(\Lambda+\rho,\varepsilon_i-\delta_1)<0$ ($\Leftrightarrow$ only if $\Lambda$
typical),\\
$\lambda_{i,m}= \lambda_{i,m+1},$ 
if $(\Lambda+\rho,\varepsilon_i-\delta_1)=0$ ($\Leftrightarrow$ only if $\Lambda$
atypical),\\
(i.e. result of Theorem \ref{pg2});
\item[(3)] For $1\leq j\leq m$, \\
$\lambda_{i+1,j+1}\geq \lambda_{i,j}\geq \lambda_{i,j+1}$ \\
(i.e. the usual $gl(j)$ branching rules);
\item[(4)] For each $r$ such that $m+1\leq r\leq m+n+1$, the $r$th row in (\ref{genGT}) must
correspond to a type 2 unitary highest $gl(m|r)$ weight, and for each $j$ such that
$1\leq j\leq m$, the $j$th row in (\ref{genGT}) must correspond to a highest $gl(j)$
weight. 
\end{itemize}
\end{thm}

\textbf{Remark}:
We may always tensor with the trivial representation $\omega(- \dot{1}|\dot{1})$ 
for $\omega \in \mathbb{R}$ to obtain $\lambda_1 = 0$ (here we have suppressed the subalgebra label since it is arbitrary). 
Noting that for atypical type 2 unitary representations there exists an even index $i$ for 
which $(\Lambda,\epsilon_1 - \epsilon_i) = 0$ we then have $\lambda_1 = \lambda_i = 0$ and 
also $\lambda_j \leq 0 $ for all $1 \leq j \leq m$ by lexicality. Furthermore, from the (a)typicality condition on the 
$gl(m|1)\supset gl(m)$ branching rule $(\Lambda+\rho,\varepsilon_i-\delta_1)\leq 0$ 
we have $\lambda_i + \lambda_{\bar{1}} + m - i \leq 0$. Again, we may set $\lambda_1 = \lambda_i = 0$ 
by tensoring with the trivial representation so that we obtain the constraint $\lambda_{\bar{1}} \leq i-m$ 
which immediately gives $\lambda_{\bar{1}} \leq 0$. Therefore it follows that $\lambda_\mu \leq 0$ for 
all odd indices $\mu$ in contrast to the covariant tensor representations for which $\lambda_i \geq 0$ 
for all even indices $i$. Here it is clear that the contravariant tensor representations which are constructed 
via tensor products of contravariant vector modules are characterized by the appearance of non-positive highest weight labels.


\section{Duality and $gl(m|1) \supset gl(m)$ branching rules}
\label{duality}
In this section we examine the consistency of the branching rules under duality.
Note that the lowering conditions on the even weights and the betweeness conditions of the odd weights are related 
under duality in the same sense that a skew Young diagram $\sigma/\nu$ that is a horizontal strip becomes a vertical
strip under conjugation (see Appendix B). 
We will now show that the additional condition on the $gl(m|1) \supset gl(m)$ branching rule in Theorem \ref{pg2} is actually essential 
to provide consistency of these lowering/betweeness conditions.

Consider an atypical type 1 unitary highest weight $\Lambda$. We may set $\lambda_n = 0$ by tensoring with the trivial 1-dimensional representation. This highest weight then takes the form 
$$
\Lambda = (\lambda_1,...,\lambda_r,\lambda_{r+1},...,\lambda_m| \omega_1,\omega_2,...,\omega_{\mu-1},0,0,...,0).
$$
where $r$ is the largest (possibly zero) even index such that $\lambda_i \geq n$ for $i \leq r$.

Note that $\mu$ immediately satisfies the second part of the atypicality condition (\ref{EqnType1Atyp}) namely $(\Lambda,\delta_\mu-\delta_n) = 0$. The first part of the atypicality condition gives
\begin{align}
(\Lambda+\rho,\varepsilon_m-\delta_\mu) &= \lambda_m + 1 - \mu \nn\\
 & = 0 \label{AtypType1FirstPart}
\end{align}
giving the modified form of highest weight
$$
\Lambda = (\lambda_1,...,\lambda_r,\lambda_{r+1},...,\mu - 1| \omega_1,\omega_2,...,\omega_{\mu-1},0,0,...,0).
$$
For a \textit{typical} type 1 unitary highest weight $\Lambda$ we necessarily have $\lambda_m \geq n$. For typical modules we
therefore set $r=m$ and $\mu = n+1$.

We now follow the method given in \cite{ZhaGou1990} to obtain the highest weight of the minimal $\mathbb{Z}$-graded component which is denoted by $\bar{\Lambda}$. From Proposition \ref{DualWeight}  we have
\begin{align}
\label{HighestWeightMinZ}
\bar{\Lambda} = \Lambda - \sum_{i=1}^m \sum_{\nu = 1}^{\mu_i} (\epsilon_i - \delta_\mu)
\end{align}
where
\begin{align*}
\mu_i &= [\mu - 1 + (\Lambda,\epsilon_i - \epsilon_m)] \wedge n \\
&= [\lambda_m + (\Lambda,\epsilon_i - \epsilon_m)] \wedge n \\
&= (\Lambda,\epsilon_i) \wedge n \\
&= \lambda_i \wedge n
\end{align*}
so that
\begin{align*}
\bar{\Lambda} &= \Lambda - \sum_{i=1}^m \sum_{\nu = 1}^{\lambda_i \wedge n} (\epsilon_i - \delta_\nu)\\
&= \Lambda - (\dot{n}_r,\dot{0} | -\dot{r} ) - \sum_{i=r+1}^m \sum_{\nu = 1}^{\lambda_i} (\epsilon_i - \delta_\nu).
\end{align*}
The weight labels of the minimal $\mathbb{Z}$-graded component $\bar{\Lambda}$ are then
\begin{align}
\bar{\lambda}_i &= \lambda_i - n,  ~~  1 \leq i \leq r \nn\\
\bar{\lambda_i} &= 0,  ~~~~ r+1 \leq i \leq m \nn\\
\bar{\lambda}_\nu &= \lambda_\nu + \# \{i| 1 \leq \lambda_i \leq \nu \}. 
\end{align}
It is then a simple procedure to obtain the highest weight of the dual module from the relation
$$
\Lambda^* = -\tau(\bar{\Lambda})
$$
where $\tau$ is the unique Weyl group element sending the positive
even roots into negative ones or, equivalently, $\tau$ has the effect of reversing the ordering of the weight labels
\begin{align*}
(\tau(\Lambda),\epsilon_i) &= (\Lambda,\epsilon_{m+1-i}) \\
(\tau(\Lambda),\delta_\nu) &= (\Lambda,\delta_{n+1-\nu}).
\end{align*}

We will now consider the $gl(m|1)$ case. The atypical type 1 unitary highest weight is now
$$
\Lambda = (\lambda_1,...,\lambda_r,0,...,0| 0)
$$
with $\lambda_i \geq 1$ for $i \leq r$ and the atypicality condition (\ref{AtypType1FirstPart}) being trivially satisfied.
The weight labels of the minimal $\mathbb{Z}$-graded component $\bar{\Lambda}$ are then
\begin{align}
\bar{\lambda}_i &= \lambda_i - 1,  ~~  1 \leq i \leq r \nn\\
\bar{\lambda_i} &= 0,  ~~~~ r+1 \leq i \leq m \nn\\
\bar{\lambda}_\nu &= \lambda_\nu + r \label{BarredLambdasGLM1}
\end{align}
which implies the weight labels of the dual module of highest weight $\Lambda^*$ are
$$
\Lambda^* = (0,...,0,1 - \lambda_r,...,1-\lambda_1| -r).
$$
The $gl(m|1) \supset gl(m)$ branching rule in Theorem \ref{pg2} states
that $\lambda_{m,m} = \lambda_{m,m+1}$. \textit{Without} this restriction, a $gl(m)$ weight such as
$$
\Lambda' = (\lambda_1,...,\lambda_r,0,..,0,-1)
$$
would be a valid $V(\Lambda) \supset V(\Lambda')$ submodule inclusion. The dual highest weight of $V(\Lambda')$ is given by
$$
\Lambda'^* =-\tau(\Lambda')
$$
so that
$$
\Lambda'^* = (1,0,...,0,-\lambda_r,...,-\lambda_1).
$$
We would then find that $V(\Lambda^*) \supset V(\Lambda'^*)$ breaks the lowering condition on the first even weight label.
Indeed we see that the type 1 $gl(m|1) \supset gl(m)$ branching rule fixes the last $m-r$ even weight labels of $\Lambda$ so that
the lowering conditions on the first $r$ even weight labels of the dual module are satisfied. Similarly, we may consider a type 2 unitary highest weight $\Lambda^*$ and set $k$ to
be the maximal even index such that $(\Lambda^*,\epsilon_k) =0$. Then the type 2 unitary $gl(m|1) \supset gl(m)$ branching rule fixes the first $k$ weight labels of $\Lambda^*$ so that
the lowering conditions on the last $m-k$ even weight labels of the dual (type 1 unitary) module are satisfied.

\section{Matrix element formulae}

\label{MEs}
We now recall some of the definitions and results from our article \cite{GIW1} which will
be used to derive the matrix element formulae of the current article. Firstly, we note that $gl(m|n+1)$ admits the subalgebra chain
$$
gl(m|n+1)\supset gl(m|n)\supset \cdots \supset gl(m|1)\supset gl(m)\supset
gl(m-1)\supset \cdots \supset gl(1).
$$
Let $p$ denote the position on the subalgebra chain so that $m+n \geq p > m$
 indicates the $gl(m|p-m)$ subalgebra while $m \geq p \geq 1$ indicates the $gl(p)$ subalgebra.
Then ${\cal A}_p$ (${\cal \bar{A}}_p$) is understood to be the vector (adjoint) matrix associated with $gl(m|p-m)$ for $p>m$ and with $gl(p)$ for $p \leq m$. 
The entries of ${\cal A}_p$ are given by
\begin{align}
	{\cal A}^{q}_{\ r} = (-1)^{(q)}E_{q r}, ~1 \leq q,r \leq p \label{equdefx}
\end{align}
and the entries of ${\cal \bar{A}}_p$ are given by
\begin{align}
{\cal \bar{A}}_q^{\ r} = - (-1)^{(q)(r)}E_{r q}, ~1 \leq q,r \leq p .\label{adjoint}
\end{align}

The associated characteristic identities are
\begin{align*}
	\prod^p_{k=1} ({\cal A}_p - \alpha_{k,p}) = 0 
\end{align*}
and
\begin{align*}
\prod_{k=1}^p ({\cal \bar{A}}_p - \bar{\alpha}_{k,p}) = 0
\end{align*}
with the characteristic roots
\begin{align}
\alpha_{k,p} &= (-1)^{(k)}(\lambda_{k,p}+m-k)-n \label{evenalpha}\\
\bar{\alpha}_{k,p} &= m-(-1)^{(k)}(\lambda_{k,p}+m+1-k)\label{oddalpha}.
\end{align}
From the characteristic identities we obtain the projections
\begin{align*}
P{p \brack r} = \prod^{p}_{k \neq r} \left( \frac{{\cal A}_p - {\alpha}_{k,p}}{{\alpha}_{r,p} - {\alpha}_{k,p}} \right)
\end{align*}
and
\begin{align*}
\bar{P}{p \brack r} = \prod^{p}_{k \neq r} \left( \frac{{\cal \bar{A}}_p - \bar{\alpha}_{k,p}}{\bar{\alpha}_{r,p}
- \bar{\alpha}_{k,p}} \right)\nn.
\end{align*}

The odd $gl(m|p)$ vector and contragredient vector operators denoted by $\psi(p)$ and $\phi(p)$ respectively are defined by
\begin{align}
\psi(p)^q &= (-1)^{(q)}E_{q,p+1}={\cal A}^q_{\ p+1}, ~1 \leq q \leq p, \label{Chap5PsiDef}\\
\phi(p)_q &= (-1)^{(q)}E_{p+1,q} = -(-1)^{(q)}{\cal \bar{A}}^{p+1}_{\ \ \ \ q}, ~1 \leq q \leq p. \label{Chap5PhiDef}
\end{align}
The vector and contragredient vector operators may be expressed as sums of shift components
\begin{align}
\psi(p)^q &= \sum^{m \wedge p}_{i=1} \psi{p \brack i}^q + \sum^{p-m}_{\mu = 1} \psi{p \brack \mu}^q,\nn\\
\phi(p)_q &= \sum^{m \wedge p}_{i=1} \phi{p \brack i}_q + \sum^{p-m}_{\mu = 1} \phi{p \brack \mu}_q,\nn
\end{align}
where $a \wedge b = min(a,b)$ and
\begin{align}
\psi{p \brack r}^p &= \psi(p)^s \bar{P}{p \brack r}_s^{\ p} = P{p \brack r}^p_{\
s}\psi(p)^s,\nn\\
\phi{p \brack r}_p &= \bar{P}{p \brack r}_p^{\ s}\phi(p)_s = (-1)^{(p)+(s)}\phi(p)_s
P{p \brack r}^s_{\ p}.\nn
\end{align}

In \cite{GIW1} we also defined the $gl(m|p)$ invariants $c_{r,p}$,~$\bar{c}_{r,p}$ where
\begin{align}
c_{r,p} &= P{p \brack r}^{p}_{\ \ \  p}  ,\nn\\
\bar{c}_{r,p} &= \bar{P}{p \brack r}_{p}^{ \ \ \ p}\label{c_sub_notation}
\end{align}
and $\delta_{r,p}$,~$\bar{\delta}_{r,p}$ which satisfy
\begin{align}
(-1)^{(p)}\psi{p \brack r}^p \phi{p \brack r}_p
&=
\delta_{r,p} P{p \brack r}^p_{\ p}
\nn\\
&= \delta_{r,p}c_{r,p}
\label{equ6.6}
\end{align}
and
\begin{align}
\phi{p \brack r}_p \psi{p \brack r}^p
=
\bar{\delta}_{r,p}\bar{c}_{r,p}.
\label{equ6.7}
\end{align}

In the case of type 2 unitary representations where
$$
\left( \psi{p \brack r}^p \right)^\dagger = (-1)^{(p)} \phi{p \brack r}_p,
$$
we note that equations (\ref{equ6.6}) and (\ref{equ6.7}) determine the square 
of the matrix elements of $\phi_{m+n}$ and $\psi_{m+n}$ respectively. 
Thus we take the formulae arising from equations (\ref{equ6.6}) and (\ref{equ6.7}) 
to determine the matrix elements. 

We now give closed form expressions for the matrix elements of the
generators $E_{l,p+1}$ and $E_{p+1,l}$ $(1 \leq l \leq p)$. Once again using the
Gelfand-Tsetlin (GT) basis notation 
with the label $\lambda_{r,p}$ located at the $r$th position in the $p$th row. 
The matrix of $E_{p+1,p+1}$ is diagonal with the entries
\begin{align}
\sum_{r=1}^{p+1} \lambda_{r,p+1} - \sum_{r=1}^{p} \lambda_{r,p}. \nn
\end{align}
We consider a fixed GT pattern denoted by $| \lambda_{q,s} \rangle$ and proceed to obtain
the matrix elements of the elementary lowering generators $E_{p+1,p}$. 

We first resolve $E_{p+1,p}$ into its shift components, which gives
\begin{align}
E_{p+1,p} | \lambda_{q,s} \rangle &= \sum_{r=1}^p (-1)^{(p)}\phi[r]_p   | \lambda_{q,s} \rangle \nn\\
&= \sum_{r=1}^p \bar{N}^p_r ( \lambda_{q,p+1} ; \lambda_{q,p} ; \lambda_{q,p-1}) 
| \lambda_{q,s} - \Delta_{r,p} \rangle, \nn
\end{align}
where $| \lambda_{q,s} - \Delta_{r,p} \rangle$ indicates the GT pattern obtained from 
$| \lambda_{q,s} \rangle$ by decreasing the label $\lambda_{r,p}$ by one unit and leaving 
the remaining labels unchanged. 

{\bf Remark:} We adopt the convention throughout the article that $| \lambda_{q,s} -
\Delta_{r,p} \rangle$ is identically zero if the branching rules are not satisfied. In
other words, $| \lambda_{q,s} - \Delta_{r,p} \rangle$ does not form an allowable GT
pattern. In such a case the matrix element is understood to be identically zero. 


Since the shift operators acting on type 2 unitary modules satisfy the Hermiticity condition
\begin{align}
\phi[r]_p = (-1)^{(p)} \left[ \psi[r]^p \right]^\dagger \nn
\end{align}
then we may use equation (\ref{equ6.6}) to express the matrix elements $\bar{N}^p_r$ as 
\begin{align}
\bar{N}^p_r ( \lambda_{q,p+1} ; \lambda_{q,p} ; \lambda_{q,p-1}) = 
\langle \lambda_{q,s} | \delta_{r,p} c_{r,p} |  \lambda_{q,s} \rangle^{1/2} \nn
\end{align}
where $\delta_{r,p}$ and $c_{r,p}$ are either invariants of the $gl(m|p-m)$ subalgebra 
for $m < p \leq m+n$ or invariants of the $gl(p)$ subalgebra for $0 < p \leq m$.

The matrix element $\bar{N}^p_r$ has an undetermined sign (or phase factor). 
However, the Baird and Biedenharn convention sets the phases of the matrix elements 
of the elementary generators $E_{p+1,p}$ to be real and positive - we will follow \cite{Gould1981} 
and adopt this convention. Matrix element phases for the non-elementary generators will be discussed later in this section.

Expressions for the eigenvalues of the invariants $c_{r,p}$ and $\delta_{r,p}$ adapted from \cite{GIW1} are 
given in terms of the characteristic roots of equations (\ref{evenalpha},\ref{oddalpha}) by 
\begin{align}
c_{i,p} &= 
\prod_{k\neq i}^m\left( \frac{\alpha_{i,p}-\alpha_{k,p}-1}{\alpha_{i,p}-\alpha_{k,p-1}} \right)
\prod_{\nu=1}^{n+1}(\alpha_{i,p}-\alpha_{\nu,p})^{-1}\prod_{\nu=1}^n(\alpha_{i,p}-\alpha_{\nu,p-1}+1),\ \ 1\leq
i\leq m,\nn\\
c_{\mu,p} &= \prod_{k=1}^m\left( \frac{\alpha_{\mu,p}-\alpha_{k,p}-1}{\alpha_{\mu,p}-\alpha_{k,p-1}} \right)
\prod_{\nu\neq \mu}^{n+1}(\alpha_{\mu,p} - \alpha_{\nu,p})^{-1} \prod_{\nu=1}^n (\alpha_{\mu,p} -
\alpha_{\nu,p-1}+1),\ \ 1\leq \mu\leq n+1.\nn
\end{align}
and
\begin{align}
\delta_{i,p} &= 
\prod_{k\neq i}^m \left( \frac{\alpha_{k,p}-\alpha_{i,p}}{\alpha_{k,p+1}-\alpha_{i,p}+1} \right)
\prod_{\nu=1}^n(\alpha_{\nu,p}-\alpha_{i,p}-1)^{-1}
\prod_{\nu=1}^{n+1}(\alpha_{\nu,p+1}-\alpha_{i,p}),\ \ 1\leq i\leq m,\nn\\
\delta_{\mu,p} &= -\prod_{k=1}^m\left(
\frac{\alpha_{k,p}-\alpha_{\mu,p}}{\alpha_{k,p+1}-\alpha_{\mu,p}+1} \right)
\prod_{\nu\neq\mu}^n(\alpha_{\nu,p}-\alpha_{\mu,p}-1)^{-1}
\prod_{\nu=1}^{n+1}(\alpha_{\nu,p+1}-\alpha_{\mu,p}),\ \ 1\leq \mu\leq n.\nn
\end{align}

For $p \geq m + 1$ we then obtain the type 2 unitary elementary lowering operator matrix elements
\begin{align}
	\bar{N}^p_i &= \Bigg[ \prod_{k\neq i = 1}^m 
	\left( 
	\frac{(\alpha_{k,p} - \alpha_{i,p})(\alpha_{k,p} - \alpha_{i,p} + 1)}
	{(\alpha_{k,p+1} - \alpha_{i,p} + 1)(\alpha_{k,p-1} - \alpha_{i,p})} 
	\right) \nn\\
	\times 
	&\left( \frac{ \prod^{p-m-1}_{\nu = 1} (\alpha_{\nu,p-1} - \alpha_{i,p} - 1)  
	\prod_{\nu=1}^{p-m+1} (\alpha_{\nu,p+1} - \alpha_{i,p})}
	{ \prod^{p-m}_{\nu \neq i = 1} (\alpha_{\nu,p} - \alpha_{i,p} )(\alpha_{\nu,p} - \alpha_{i,p} - 1 )}
	\right) \Bigg]^{1/2}, ~~p \geq m+1 \label{SimpleLoweringEven} \\
\bar{N}^p_\mu &= \Bigg[ \prod_{k = 1}^m 
\left( \frac{(\alpha_{k,p} - \alpha_{\mu,p} + 1)(\alpha_{k,p} - \alpha_{\mu,p} )}
{(\alpha_{k,p-1} - \alpha_{\mu,p})(\alpha_{k,p+1} - \alpha_{\mu,p} + 1)} 
\right) \nn\\
	\times 
	&\left( \frac{ \prod^{p-m-1}_{\nu = 1} (\alpha_{\nu,p-1} - \alpha_{\mu,p} - 1)  
	\prod_{\nu=1}^{p-m+1} (\alpha_{\nu,p+1} - \alpha_{\mu,p})}
	{ \prod^{p-m}_{\nu \neq \mu = 1} (\alpha_{\nu,p} - \alpha_{\mu,p} )(\alpha_{\nu,p} - \alpha_{\mu,p} - 1 )} 
	\right) \Bigg]^{1/2}, ~~p \geq m+1. \label{SimpleLoweringOdd}
\end{align}

We may now obtain matrix elements of the \textit{raising} operators $E_{p,p+1}$ via the relation
\begin{align}
\langle \lambda_{q,s} + \Delta_{r,p}| E_{p,p+1} |\lambda_{q,s}\rangle  
= \langle \lambda_{q,s} | E_{p+1,p} |\lambda_{q,s} + \Delta_{r,p}\rangle  \nn,
\end{align}
which holds on type 2 unitary representations.
It is clear that, $E_{p,p+1}$ is simply obtained 
from  $E_{p+1,p}$ by making the substitution $\lambda_{rp} \rightarrow \lambda_{rp} + 1$ within the 
characteristic roots occurring in the matrix element formula for $E_{p+1,p}$. 
From equations (\ref{evenalpha}) and (\ref{oddalpha}) we see this shift of the 
label $\lambda_{rp}$ is equivalent to the substitutions
$$
\alpha_{i,p} \rightarrow \alpha_{i,p} + 1,~~~\alpha_{\mu,p,} \rightarrow \alpha_{\mu,p} - 1.
$$

After applying the above substitutions to the matrix element equations (\ref{SimpleLoweringEven}) and (\ref{SimpleLoweringOdd}) we then have
 the elementary raising generator matrix elements \begin{align}
	N^p_i &= \Bigg[ \prod_{k\neq i = 1}^m 
	\left( \frac{(\alpha_{k,p} - \alpha_{i,p} - 1)(\alpha_{k,p} - \alpha_{i,p})}
	{(\alpha_{k,p+1} - \alpha_{i,p})(\alpha_{k,p-1} - \alpha_{i,p} - 1)} \right) \nn\\
	\times 
	&\left( \frac{ \prod^{p-m-1}_{\nu = 1} (\alpha_{\nu,p-1} - \alpha_{i,p} - 2)  
	\prod_{\nu=1}^{p-m+1} (\alpha_{\nu,p+1} - \alpha_{i,p} - 1)}
	{ \prod^{p-m}_{\nu \neq i = 1} (\alpha_{\nu,p} - \alpha_{i,p} - 1)(\alpha_{\nu,p} - \alpha_{i,p} - 2 )} 
	\right) \Bigg]^{1/2}, ~~p \geq m+1 \label{SimpleRaisingEven} \\
N^p_\mu &= \Bigg[ \prod_{k = 1}^m 
\left( \frac{(\alpha_{k,p} - \alpha_{\mu,p} + 2)(\alpha_{k,p} - \alpha_{\mu,p} + 1 )}
{(\alpha_{k,p-1} - \alpha_{\mu,p} + 1)(\alpha_{k,p+1} - \alpha_{\mu,p} + 2)} 
\right) \nn\\
	\times 
	&\left( \frac{ \prod^{p-m-1}_{\nu = 1} (\alpha_{\nu,p-1} - \alpha_{\mu,p} ) 
	 \prod_{\nu=1}^{p-m+1} (\alpha_{\nu,p+1} - \alpha_{\mu,p} + 1)}
	 { \prod^{p-m}_{\nu \neq \mu = 1} (\alpha_{\nu,p} - \alpha_{\mu,p} + 1 )(\alpha_{\nu,p} - \alpha_{\mu,p} )} 
	 \right) \Bigg]^{1/2}, ~~p \geq m+1. \label{SimpleRaisingOdd}
\end{align}

\textbf{Remark}: We observe that the above type 2 unitary matrix element equations for $p \geq m+1$ match the  
type 1 unitary matrix element equations given in \cite{GIW2} (page 17). 
Using the same procedure as above it may be shown that the $p=m$ type 2 unitary matrix 
element equations (given below) also match the $p=m$ type 1 unitary matrix element equations.
Finally, the $p<m$ case is given by the $gl(m)$ matrix element results of \cite{Gould1981}. 
It is important to note that the branching rules 
and therefore the vanishing conditions of the matrix elements are different between the two representation types. Furthermore, for the non-elementary generators, there is a difference of phase that will be given later in this section.

For the $p=m$ case we have
\begin{equation}
\bar{N}^m_i =  (\alpha_{m+1,m+1}-\alpha_{i,m})^{1/2} 
\left( \frac{\prod_{k}^{m-1} (\alpha_{k,m-1} - \alpha_{i,m} + 1)}
{ \prod_{k\neq i}^m (\alpha_{k,m+1}-\alpha_{i,m} + 1)} \right)^{1/2} \nn  
\end{equation}
which after the substitution $\alpha_{i,p} \rightarrow \alpha_{i,p} + 1$ gives
\begin{equation}
N^m_i =   (\alpha_{m+1,m+1}-\alpha_{i,m}-1)^{1/2} \left( \frac{\prod_{k}^{m-1} (\alpha_{k,m-1} - \alpha_{i,m})}{ \prod_{k\neq
i}^m (\alpha_{k,m+1}-\alpha_{i,m} ) } \right)^{1/2} \nn 
\end{equation}

Finally, when $p<m$, we make use of the results in \cite{Gould1981}, namely
\begin{align}
	\bar{N}^p_i = \left( \frac{(-1)^p \prod_{k = 1}^{p+1} (\alpha_{k,p+1} - \alpha_{i,p}) \prod_{k=1}^{p-1} (\alpha_{i,p} - \alpha_{k,p-1} - 1)}{\prod_{k \neq i}^p (\alpha_{i,p} -
\alpha_{k,p} ) (\alpha_{i,p} - \alpha_{k,p} - 1)  } \right)^{1/2} , \ \ p < m \nn
\end{align}
and
\begin{align}
	N^p_i =  \left( \frac{(-1)^p \prod_{k = 1}^{p+1} (\alpha_{k,p+1} - \alpha_{i,p} -
1) \prod_{k=1}^{p-1} (\alpha_{i,p} - \alpha_{k,p-1})}{\prod_{k \neq i}^p (\alpha_{i,p} -
\alpha_{k,p} + 1) (\alpha_{i,p} - \alpha_{k,p})  } \right)^{1/2} , \ \ p < m.\nn
\end{align}

We now turn to the non-elementary generators $E_{p+1,l}$ and $E_{l,p+1}$. Resolving the  $E_{p+1,l}$ $(l < p)$ into simultaneous shift components, we have
\begin{align}
E_{p+1,l} | \lambda_{q,s} \rangle &= \sum_u \bar{N}{p~\ldots~l \brack u_{p} \ldots u_l}  
| \lambda_{q,s} - \Delta_{u_p,p} -\ldots - \Delta_{u_l,l} \rangle, \nn
\end{align}
where $| \lambda_{q,s} - \Delta_{u_p,p} - \ldots - \Delta_{u_l,l}  \rangle$ indicates the GT pattern 
obtained from $| \lambda_{q,s} \rangle$ by decreasing the $p-l+1$ labels $\lambda_{u_r,r}$ of the 
subalgebra $gl(m|r-m)$ for $r=l,\ldots,p$, by one unit and leaving the remaining labels unchanged, and the summation symbol is shorthand notation for
$$
\sum_{u_{m+n} = 1}^{m+n} \sum_{u_{m+n-1} = 1}^{m+n-1} \cdots \sum_{u_p = 1}^p.
$$
Similarly, we also have 
$$
E_{l,p+1} | \lambda_{q,s} \rangle = \sum_u N{p~\ldots~l \brack u_{p} \ldots u_l} 
| \lambda_{q,s} + \Delta_{u_p,p} + \ldots + \Delta_{u_l,l} \rangle.
$$

The matrix elements of these non-elementary generators also match those of the type 1 unitary case. By following the
derivation given in \cite{GIW2} we obtain
\begin{align}
\bar{N} {p~\ldots~l \brack u_{p} \ldots u_l} 
=\frac{S \left(\bar{N}{p~\ldots~l \brack u_{p} \ldots u_l} \right) \prod_{r=l}^p \bar{N}^{r}_{u_r}}{
\prod_{s=l+1}^p  \sqrt{(\alpha_{u_s,s}-\alpha_{u_{s-1},s-1} + 1)
(\alpha_{u_s,s}-\alpha_{u_{s-1},s-1}) } }, \label{Type2_NonEle_Type2}
\end{align}

\begin{align}
N{p~\ldots~l \brack u_{p} \ldots u_l}
= \frac{S \left(N{p~\ldots~l \brack u_{p} \ldots u_l} \right) \prod_{r=l}^p N^{r}_{u_r}}{
\prod_{s=l+1}^p  \sqrt{(\bar{\alpha}_{u_s,s}-\bar{\alpha}_{u_{s-1},s-1} + 1)
(\bar{\alpha}_{u_s,s}-\bar{\alpha}_{u_{s-1},s-1}) } }, \label{Type2_NonEle}
\end{align}
where the signs of the type 2 unitary matrix elements are given by the expression
\begin{align}
S\left(\bar{N}{p~\ldots~l \brack u_{p} \ldots u_l}\right) &=S\left(N{p~\ldots~l \brack u_{p} \ldots u_l}\right) \nn\\
&= \prod^p_{s=l+1} (-1)^{(s+1)}(-1)^{(u_{s-1})(u_s)+(u_{s-1})+(u_s)} S(u_s - u_{s-1})
\label{phase} 
\end{align}
and where $S(x) \in \{-1,1\}$ is the sign of $x$, $S(0) = 1$ and, as usual, odd indices are
considered greater than even indices. Details of the phase calculation are given in Appendix A.

%

{\bf Remarks: } \label{importantremarks}
\begin{enumerate}
\item
It is understood that to apply the matrix element formula derived above, where possible
terms are canceled first and reduced to the most simplified rational form before
applying the formulae and substituting weight labels.
\item
All terms appearing in the square roots in the above formula are indeed positive numbers.
\item
We remind the reader that in all cases we have adopted the convention that a shifted
pattern $| \lambda_{q,s} \pm \Delta_{r,p} \rangle$ is identically zero if the branching rules are not satisfied. 
\end{enumerate}

We would like to emphasize the surprising nature of the correspondence between the type 1 unitary and the type 2 unitary matrix element equations. In short, the caution exercised to ensure that we always tensor the (type 1 unitary) vector module $V(\varepsilon_1)$ with a type 1 unitary module $V(\Lambda)$, while being technically essential to obtain complete reducibility, was actually inessential in obtaining the resulting matrix element expression. However the vanishing conditions and phases of the matrix elements are dependent on the type of the module concerned. Therefore the general procedure to find matrices of generators of $gl(m|n+1)$
(including non-elementary ones) corresponding to a type 1 or type 2 unitary irreducible highest
weight module $V(\Lambda)$ is:
\begin{itemize}
\item[1.] Identify the type ($1$ or $2$) of the highest weight $\Lambda$ using the classification results of Theorems \ref{gzthm1} and \ref{gzthm2};
\item[2.] Determine the branching rules for the whole subalgebra chain
(\ref{flag}), using Theorem \ref{mainbranchingrule} for type 2 representations or Theorem 9 within \cite{GIW2} for type 1 representations;
\item[3.]
Express every basis vector as a GT pattern of the form (\ref{genGT});
\item[4.] 
Determine the matrix elements using the formulae presented in Section \ref{MEs}.
\end{itemize}

\section*{Acknowledgments}
This work was supported by the Australian Research Council through Discovery Project
DP140101492. J.L.W. acknowledges the support of an Australian Postgraduate Award.
%
\section*{Appendix A: Phase convention}

\label{Type2Phases}
We will now derive the phase of the matrix elements of the generators $E_{p+2,p}$ and then extend this 
result to matrix elements of all generators $E_{p+2,p-q}$. This calculation is analogous to the type 1 unitary 
case given in \cite{GIW2} but care must be taken when shifting two labels of differing parity.

The simple generators $E_{p+1,p}$ acting on a GT pattern $|\lambda_{q,s} \rangle$
(with the top row being the highest weight of a type 2 unitary representation for $gl(m|n+1)$) will produce
\begin{align} 
E_{p+1,p} |\Lambda \rangle = \sum^p_{a=1} {\bar{N}}^{p}_a |\Lambda - \varepsilon_{a,p} \rangle \nn
\end{align}
where $|\lambda_{q,s} - \varepsilon_{a,p} \rangle$ is the GT pattern $|\lambda_{q,s} \rangle$ but with
the $a$th label of the $p$th row shifted by $-1$.
Consequently, non-zero matrix elements of the simple generators will be of the form
\begin{align}
\langle \lambda_{q,s} - \varepsilon_{a,p} | E_{p+1,p} |\lambda_{q,s} \rangle  = + {\bar{N}}^{p}_a [\lambda_{q,s}] \label{BasicME2}
\end{align}
where we have set ${\bar{N}}^{p}_a$ to be positive by the Baird and Biedenharn convention. 
Non-zero matrix elements of non-simple generators $E_{p+2,p}$ are given by
\begin{align}
{\bar{N}}^{p+1~p}_{~b~~~a} &= \langle \lambda_{q,s} - \varepsilon_{a,p} - \varepsilon_{b,p+1} | E_{p+2,p} | \lambda_{q,s} \rangle \nn\\
&= \langle \lambda_{q,s} - \varepsilon_{a,p} - \varepsilon_{b,p+1} | [E_{p+2,p+1} , E_{p+1,p}] | \lambda_{q,s} \rangle \nn\\ 
&=  \langle \lambda_{q,s} - \varepsilon_{a,p} - \varepsilon_{b,p+1} | E_{p+2,p+1} | \lambda_{q,s} - \varepsilon_{a,p} \rangle \langle \lambda_{q,s} - \varepsilon_{a,p} | E_{p+1,p}  | \lambda_{q,s} \rangle. \nn\\
&~- \langle \lambda_{q,s} - \varepsilon_{a,p} - \varepsilon_{b,p+1} | E_{p+1,p} | \lambda_{q,s} - \varepsilon_{b,p+1} \rangle \langle \lambda_{q,s} - \varepsilon_{b,p+1} | E_{p+2,p+1} | \lambda_{q,s} \rangle. \nn
\end{align}
Using (\ref{BasicME2}) the above equation can be written as 
\begin{align}
{\bar{N}}^{p+1~p}_{~b~~~a} = {\bar{N}}^{p+1}_b [\lambda_{q,s} - \varepsilon_{a,p}] {\bar{N}}^{p}_a [\lambda_{q,s}] - {\bar{N}}^{p}_a [\lambda_{q,s} - \varepsilon_{b,p+1} ] {\bar{N}}^{p+1}_b [\lambda_{q,s}] \nn
\end{align}
where all of the matrix elements on the RHS are positive due to the Baird-Beidenharn convention.

Recall the following formulae together with the definitions of $I$ and $\tilde{I}$ given in \cite{GIW1}:
\begin{align}
\label{ToShift1}
\delta_{a,p} &=  \prod_{k\in I,k\neq a} \left(\alpha_{a,p} - \alpha_{k,p} -
(-1)^{(k)}\right)^{-1}\prod_{r\in\tilde{I}} \left(\alpha_{a,p} - \alpha_{r,p+1} \right),\ \ a\in I',
\end{align}
\begin{align}
\label{ToShift2}
c_{b,p+1} = \prod_{k\in\tilde{I},k\neq b} (\alpha_{b,p+1} - \alpha_{k,p+1})^{-1}\prod_{r\in I}(\alpha_{b,p+1} -
\alpha_{r,p}-(-1)^{(r)}),\ \ b\in \tilde{I}.
\end{align}
By examining the change (appearing as the addition or removal of terms) resulting from the shift $\lambda_{q,s} - \varepsilon_{b,p+1}$ to equation (\ref{ToShift1}) and
the shift $\lambda_{q,s} - \varepsilon_{a,p}$ to equation (\ref{ToShift2}) we find that for odd $a$ and odd $b$ :
\begin{align*}
{\bar{N}}^{p+1~p}_{~b~~~a} [\Lambda] &= \bar{N}^{p+1}_b [\lambda_{q,s} - \varepsilon_{a,p}] \bar{N}^{p}_a [\lambda_{q,s}] - \bar{N}^{p}_a [\lambda_{q,s} - \varepsilon_{b,p+1} ] \bar{N}^{p+1}_b [\lambda_{q,s}] \nn\\
&= ({\delta}_{b,p+1} {c}_{b,p+1})^{1/2} [\lambda_{q,s} - \varepsilon_{a,p}] \bar{N}^{p}_a [\lambda_{q,s}] - ({\delta}_{a,p} {c}_{a,p})^{1/2} [\lambda_{q,s} - \varepsilon_{b,p+1} ] \bar{N}^{p+1}_b [\lambda_{q,s}] \nn\\
&=  ({c}_{b,p+1})^{1/2} [\lambda_{q,s} - \varepsilon_{a,p}] (\bar{\delta}_{b,p+1})^{1/2} [\lambda_{q,s}] \bar{N}^{p}_a [\lambda_{q,s}] \nn\\
&~~- ({\delta}_{a,p})^{1/2} [\lambda_{q,s} - \varepsilon_{b,p+1} ] ({c}_{a,p})^{1/2} [\lambda_{q,s}]  \bar{N}^{p+1}_b [\lambda_{q,s}]  \nn\\
&= \frac{\sqrt{{\alpha}_{b,p+1} - {\alpha}_{a,p}}}{\sqrt{{\alpha}_{b,p+1} - {\alpha}_{a,p} + 1} }  ({c}_{b,p+1})^{1/2} ({\delta}_{b,p+1})^{1/2} [\lambda_{q,s}] \bar{N}^p_a [\lambda_{q,s}]  \\
&~~- \frac{ \sqrt{{\alpha}_{a,p} - {\alpha}_{b,p+1} - 1} }{ \sqrt{{\alpha}_{a,p} - {\alpha}_{b,p+1}} } ({\delta}_{a,p})^{1/2} ({c}_{a,p})^{1/2} [\lambda_{q,s}]  \bar{N}^{p+1}_b [\lambda_{q,s}]   \nn\\
&= \left( \sqrt{\frac{{\alpha}_{a,p} - {\alpha}_{b,p+1} }{{\alpha}_{a,p} - {\alpha}_{b,p+1} - 1} } - \sqrt{\frac{ {\alpha}_{a,p} - {\alpha}_{b,p+1} - 1 }{ {\alpha}_{a,p} - {\alpha}_{b,p+1} }} \right)  \bar{N}^{p}_a \bar{N}^{p+1}_b  [\lambda_{q,s}] \nn\\
&= ( {\alpha}_{a,p} - {\alpha}_{b,p+1} - 1)^{-1/2} ({\alpha}_{a,p} - {\alpha}_{b,p+1})^{-1/2} \bar{N}^{p}_a \bar{N}^{p+1}_b  [\lambda_{q,s}]. \nn
\end{align*} 
Similarly, for the cases
corresponding to the other three parity combinations of $a$ and $b$, we obtain the same result.

We observe that the sign of ${\bar{N}}^{p~p+1}_{~a~b}$ is directly given by the sign of
${\alpha}_{a,p} - {\alpha}_{b,p+1}$. However, we must also note that in the $gl(m)$ case where $p<m$  
 the sign of ${\bar{N}}^{p~p+1}_{~a~b}$ is given by the sign of ${\alpha}_{b,p+1} - {\alpha}_{a,p}$ \cite{Gould1981}.  Furthermore, it was shown in \cite{GIW2} that the overall 
sign of $\bar{N}{p~\ldots~l \brack u_{p} \ldots u_l}$ is given by the multiplied signs of such terms at each level of the 
subalgebra chain as follows 
\begin{align}
S \left(\bar{N}{p~\ldots~l \brack u_{p} \ldots u_l} \right) = \prod^p_{s=l+1} (-1)^{(s+1)} S({\alpha}_{u_{s-1},s-1} - {\alpha}_{u_s,s}). \label{ProductOfPhases2}
\end{align}
where we have added the $(-1)^{(s+1)}$ grading factor to include the $gl(m)$ case.

Now, for $(u_s) = 0$, $(u_{s-1})=0$,$u_s \neq u_{s-1}$ we have
\begin{align}
S({\alpha}_{u_{s-1},s-1} - {\alpha}_{u_s,s}) &= S(\lambda_{u_{s-1},s-1} - \lambda_{u_s,s}  + u_s - u_{s-1}) \nn\\
&= S(u_s-u_{s-1})
\end{align}
by lexicality. 

For $(u_s) = 1$, $(u_{s-1})=1$ we have
\begin{align}
S({\alpha}_{u_{s-1},s-1} - {\alpha}_{u_s,s}) &= S(\lambda_{u_s,s} - \lambda_{u_{s-1},s-1} + u_{s-1} - u_s) \nn\\
&= S(u_{s-1}-u_s). \nn
\end{align}

For $(u_s) = 1$, $(u_{s-1})=0$
\begin{align}
S({\alpha}_{u_{s-1},s-1} - {\alpha}_{u_s,s}) &= S({\alpha}_{u_{s-1},s} - {\alpha}_{u_s,s} - 1) \nn\\
&= S(~(\Lambda_{(s)} + \rho_{(s)}, \epsilon_{u_{s-1}} - \delta_{u_s} )~-2) \nn
\end{align}
where we have denoted $\Lambda_{(p)}$ and $\rho_{(p)}$ to be the highest weight and graded half-sum of the positive roots restricted to the subalgebra level $gl(m|p-m)$. For $\Lambda$ typical type 2 unitary we have $(\Lambda_{(s)} + \rho_{(s)},\epsilon_1 - \delta_1) < 0$ which gives
\begin{align}
(\Lambda_{(s)} + \rho_{(s)}, \epsilon_{u_{s-1}} - \delta_{u_s} ) &= (\Lambda_{(s)} + \rho_{(s)}, \epsilon_1 -
\delta_1 ) + (\Lambda_{(s)} + \rho_{(s)}, \epsilon_{u_{s-1}} - \epsilon_1 ) + (\Lambda_{(s)} + \rho_{(s)},
\delta_1 - \delta_{u_s}) \nn\\
&  \leq (\Lambda_{(s)} + \rho_{(s)},\epsilon_1 - \delta_1) < 0 \nn
\end{align} 
where we have used the fact that $\Lambda_{(s)} + \rho_{(s)} \in D^+$.
For $\Lambda$ atypical type 2 unitary there exists an even index $1 \leq k \leq m$ such that $(\Lambda_{(s)} + \rho_{(s)},\epsilon_k - \delta_1) = 0$ and $(\Lambda_{(s)},\epsilon_k - \epsilon_1) = 0$. Since the labels $\lambda_{j,s}$ for $1 \leq j \leq k$ are all equal, only even labels $\lambda_{u_{s-1},s}$ for $u_{s-1} \geq k$ may be lowered. For this matrix element we necessarily have $u_{s-1} \geq k$ giving
\begin{align}
(\Lambda_{(s)} + \rho_{(s)}, \epsilon_{u_{s-1}} - \delta_{u_s} ) &= (\Lambda_{(s)} + \rho_{(s)}, \epsilon_k - \delta_1 ) + (\Lambda_{(s)} + \rho_{(s)}, \epsilon_{u_{s-1}} - \epsilon_k ) + (\Lambda_{(s)} + \rho_{(s)}, \delta_1 - \delta_{u_s}) \nn\\
&= (\Lambda_{(s)} + \rho_{(s)}, \epsilon_{u_{s-1}} - \epsilon_k ) + (\Lambda_{(s)} + \rho_{(s)}, \delta_1 - \delta_{u_s}) \leq 0, \nn
\end{align}
which shows that for this case the matrix element is negative, i.e.
\begin{align}
S({\alpha}_{u_{s-1},s-1} - {\alpha}_{u_s,s}) &= -1,~~~~(u_s) = 1,(u_{s-1})=0, \nn
\end{align}
and similarly
\begin{align}
S({\alpha}_{u_{s-1},s-1} - {\alpha}_{u_s,s}) &= 1,~~~~(u_s) = 0,(u_{s-1})=1. \nn
\end{align}
Combining the above four cases gives
\begin{align}
S({\alpha}_{u_{s-1},s-1} - {\alpha}_{u_s,s}) =  (-1)^{(u_{s-1})(u_s) + (u_{s-1}) + (u_s)} S(u_s - u_{s-1}) \nn
\end{align}
where, as usual, odd indices are considered greater than even indices. Finally, from equation (\ref{ProductOfPhases2}) we have the result
\begin{align}
S\left({\bar{N}}{p~\ldots~l \brack u_{p} \ldots u_l}\right) &= \prod^p_{s=l+1} (-1)^{(s+1)} (-1)^{(u_{s-1})(u_s) + (u_{s-1}) + (u_s)} S(u_s - u_{s-1}) \label{Type2Phase_Lowering}
\end{align}
where the grading factor (for odd $s$) is the same sign as the type 1 unitary case when $(u_{s-1}) = (u_s)$ and the opposite sign when $(u_{s-1}) \neq (u_s)$. Analogously we also have
\begin{align}
S\left({{N}}{p~\ldots~l \brack u_{p} \ldots u_l}\right) &= \prod^p_{s=l+1} (-1)^{(s+1)} (-1)^{(u_{s-1})(u_s) + (u_{s-1}) + (u_s)} S(u_s - u_{s-1}) \label{Type2Phase_Raising}
\end{align}
so that for a type $\theta$ representation $(\theta \in \{1,2\})$
\begin{align}
S\left({{N}}{p~\ldots~l \brack u_{p} \ldots u_l}\right) &= S\left({\bar{N}}{p~\ldots~l \brack u_{p} \ldots u_l}\right) = \prod^p_{s=l+1}  (-1)^{(s+1)} (-1)^{(u_{s-1})(u_s) + (\theta-1)[(u_{s-1}) + (u_s)]} S(u_s - u_{s-1}). \label{TypeAnyPhase}
\end{align}

\textbf{Remark}: It is interesting to note (from closer analysis of the above proof) that the phases of the non-zero matrix elements
$$
{\bar{N}}{p~p-1 \brack u_{p} ~u_{p-1}},~~N{p~p-1 \brack u_{p} ~u_{p-1}}, ~~p > m
$$
for both type 1 unitary and type 2 unitary modules are ultimately given by the sign of
$$
(\Lambda + \rho, \alpha), ~~\alpha \in \Phi_0 \cup \Phi_1
$$
where 
\begin{align*}
\alpha &= \epsilon_{u_{p-1}} - \epsilon_{u_{p}},  ~~(u_{p-1}) = (u_p) = 0 \\
\alpha &= \delta_{u_{p-1}} - \delta_{u_{p}},  ~~(u_{p-1}) = (u_p) = 1 \\
\alpha &= \epsilon_{u_{p-1}} - \delta_{u_{p}},  ~~(u_{p-1}) = 0, (u_p) = 1 \\
\alpha &= \delta_{u_{p-1}} - \epsilon_{u_{p}},  ~~(u_{p-1}) = 1, (u_p) = 0 .
\end{align*}

From this point we may obtain the final phase expression by assuming
$$
u_{p-1} < u_p
$$
so that $\alpha \in \Phi^+_0$ for $(u_{p-1}) = (u_p)$ and $\alpha \in \Phi^+_1$ for $(u_{p-1})=0, (u_p)=1$. We may later remove the restriction $u_{p-1} < u_p$ by swapping labels to obtain the opposite sign.

When $\alpha \in \Phi^+_0$ the sign of $(\Lambda + \rho, \alpha)$ is positive for $(u_{p-1}) = (u_p) = 0$ and negative for $(u_{p-1}) = (u_p) = 1$ since both $\Lambda$ and $\rho$ are lexical. Note that this holds for both type 1 unitary and type 2 unitary $\Lambda$.

We now consider the case $\alpha \in \Phi^+_1$. The expression $(\Lambda + \rho, \alpha)$ is given by
$$
(\Lambda + \rho, \epsilon_{u_{p-1}} - \delta_{u_p}).
$$
From the previous calculations in this appendix we see that for type 1 unitary $\Lambda$ this expression is \textit{positive} while for type 2 unitary $\Lambda$ this expression is \textit{negative}. Note that this strong result is only possible due to the restrictions on the values of $u_{p-1}$ and $u_p$ for non-vanishing matrix elements.
Therefore, for type 1 unitary $\Lambda$ we can give the sign of $(\Lambda + \rho, \alpha)$ as
$$
(-1)^{(u_{p-1})(u_p)}  S(u_p - u_{p-1})
$$
and for type 2 unitary $\Lambda$ the sign is 
$$
(-1)^{(u_{p-1})(u_p) + (u_{p-1}) + (u_p)}  S(u_p - u_{p-1})
$$

\section*{Appendix B: Duality of betweeness conditions}

In this appendix we investigate the $gl(m|n)$ dual branching rules for $n>1$ via Young diagram methods.

In section \ref{duality} the form of a type 1 unitary highest weight was given as 
$$
\Lambda = (\lambda_1,...,\lambda_r,\lambda_{r+1},...,\mu - 1| \omega_1,\omega_2,...,\omega_{\mu-1},0,0,...,0).
$$
where $r$ is the largest (possibly zero) even index such that $\lambda_i \geq n$ for $i \leq r$ and where $\mu$ 
satisfies the second part of the atypicality condition (\ref{EqnType1Atyp}). We also noted that for 
typical type 1 unitary modules we necessarily have $\lambda_m \geq n$ and therefore set  $r=m$ and $\mu = n+1$ in that case.

The weight labels of the minimal $\mathbb{Z}$-graded component $\bar{\Lambda}$ were then found to be
\begin{align}
\bar{\lambda}_i &= \lambda_i - n,  ~~  1 \leq i \leq r \nn\\
\bar{\lambda_i} &= 0,  ~~~~ r+1 \leq i \leq m \nn\\
\bar{\lambda}_\nu &= \lambda_\nu + \# \{i| 1 \leq \lambda_i \leq \nu \} \label{BarredLambdas}
\end{align}
where $\#$ denotes the cardinality of the given set. The highest weight of the dual module is then
\begin{equation}
\label{DualFromZGC}
\Lambda^* = -\tau(\bar{\Lambda})
\end{equation}
where $\tau$ is the unique Weyl group element sending the positive
even roots into negative ones.

The method of obtaining $\bar{\Lambda}$ given by equation (\ref{HighestWeightMinZ}) can be expressed using Young diagrammatic methods by
considering equation (\ref{BarredLambdas}).

Let $\sigma_0$ be the partition (or equivalently the corresponding Young diagram) given by the even weights of $\Lambda$
$$
\sigma_0 = (\lambda_1,...,\lambda_m)
$$
and similarly let $\sigma_1$ be the partition (Young diagram) given by then odd weights of $\Lambda$
$$
\sigma_1 = (\lambda_{m+1},...,\lambda_{m+n}).
$$
so that we have the bipartition denoted by
$$
\sigma = (\sigma_0,\sigma_1).
$$

We now restrict to the case $r=0$ and $\lambda_\nu = 0, \forall \nu$ in equation (\ref{BarredLambdas}). We then see that 
the sequence of odd weight labels ${\bar{\lambda}_\nu}$ is precisely the conjugate partition of the sequence of even weight labels ${\lambda_i}$.
For this restricted case, we can therefore express equation (\ref{BarredLambdas}) in terms of Young diagrams as
\begin{align}
\label{YDbar}
\bar{\sigma} = (\emptyset,\sigma'_0)
\end{align}
where $\emptyset$ represents the empty partition and the superscripted prime denotes the conjugate partition. 


Equation (\ref{DualFromZGC}) expressed in terms of a Young diagram $\bar{\sigma}$ is a just a reversal of the original diagram's row ordering followed by a reflection across the vertical axis to represent negative values. The resulting diagram $\sigma^*$ is therefore, for our purposes, equivalent to the original diagram of the highest $\mathbb{Z}$-graded component $\bar{\sigma}$.

We will now give the $gl(m|n)$ branching rule for $n>1$ in terms of Young diagrams. For two partitions $\sigma$ and $\upsilon$ with $\sigma_i \geq \upsilon_i$ we denote the skew Young diagram as $\sigma/\upsilon$ as the one obtained by removing the diagram of $\upsilon$ from the diagram of $\sigma$.
A skew diagram is called a horizontal strip (vertical strip) if each
column (row) of the skew diagram contains exactly one box. We may reexpress the $gl(m|n), n>1$ branching rule (Theorem \ref{pg1}) as follows

\begin{thm}
\label{BRstrip}
Let $\sigma = (\sigma_0,\sigma_1)$ and $\upsilon = (\upsilon_0,\upsilon_1)$ be given by the bipartitions corresponding to rows $m+k+1$ and $m+k$ of a GT pattern. Then the bipartitions $\sigma$ and $\upsilon$ must satisfy the conditions
\begin{itemize}
\item[(1)] $\sigma_0/\upsilon_0$ is a horizontal strip
\item[(2)] $\sigma_1/\upsilon_1$ is a vertical strip.
\end{itemize}
\end{thm}
The above expression for the branching rule is related to the branching rule derived in \cite{ClarkPeng2015}. However, our branching rule here has been derived algebraically and applies to both covariant and contravariant tensor representations while the result given in \cite{ClarkPeng2015} has been arrived at via diagrammatic methods that apply only to covariant tensor representations (albeit for a general Borel subalgebra while we use the standard Borel).

Obviously, if a skew Young diagram is a horizontal (vertical) strip then the conjugate skew Young diagram is a vertical (horizontal) strip. Hence, from equation (\ref{YDbar}) the dual branching rule (for the restricted case under consideration) is given by 

\begin{thm}
Fix the top row of a GT pattern to be a $gl(m,n+1)$ highest weight of an atypical type 1 unitary module with even labels $\lambda_i \leq n, 1 \leq i \leq m$ and all odd labels zero. Let $\sigma = (\sigma_0,\sigma_1)$ and $\upsilon = (\upsilon_0,\upsilon_1)$ be given by the bipartitions corresponding to rows $m+k+1$ and $m+k$ with $1 \geq k \geq n$ of the GT pattern. Then the bipartitions $\sigma^*$ and $\upsilon^*$ of the corresponding rows of the dual GT pattern must satisfy 
$$
\sigma^*_1/\upsilon^*_1 = \sigma'_0/\upsilon'_0 ~~\hbox{is a vertical strip}
$$
\end{thm}

By comparing the above theorem with Theorem \ref{BRstrip} we see that the lowering conditions on the even weight labels are \textit{dual} to the $gl(n)$ betweeness conditions on the odd weight labels.


%
%
%

\end{document}